\documentclass[runningheads]{llncs}
\usepackage{hyperref}
\usepackage[T1]{fontenc}
\usepackage{amsfonts}
\usepackage{graphicx}
\usepackage{amsmath}
\usepackage[table,xcdraw]{xcolor}
\usepackage{color}

\newcommand{\subpara}[1]{\vspace{0.5em} \noindent \textbf{#1}}
\begin{document}
\title{Multi-Head Graph Convolutional Network for Structural Connectome Classification}
\titlerunning{Multi-Head GCN for Structural Connectome Classification}
\author{Anees Kazi\inst{1,2} \and
 Jocelyn Mora\inst{1} \and
 Bruce Fischl\inst{1,2} \and
 Adrian V. Dalca\inst{1,2,3} \and
 Iman Aganj \inst{1,2} }
\authorrunning{A. Kazi et al.}
%
\institute{Athinoula A. Martinos Center for Biomedical Imaging, Radiology Department, Massachusetts General Hospital, Boston, USA \and Radiology Department, Harvard Medical School, Boston, USA \and
CSAIL, Massachusetts Institute of Technology, Cambridge, USA \\
\email{akazi1@mgh.harvard.edu}}
%
\maketitle              
\begin{abstract}
We tackle classification based on brain connectivity derived from diffusion magnetic resonance images. We propose a machine-learning model inspired by graph convolutional networks (GCNs), which takes a brain-connectivity input graph and processes the data separately through a parallel GCN mechanism with multiple heads. 
The proposed network is a simple design that employs different heads involving graph convolutions focused on edges and nodes, thoroughly capturing representations from the input data. 
To test the ability of our model to extract complementary and representative features from brain connectivity data, we chose the task of sex classification. This quantifies the degree to which the connectome varies depending on the sex, which is important for improving our understanding of health and disease in both sexes. We show experiments on two publicly available datasets: PREVENT-AD (347 subjects) and OASIS3 (771 subjects). The proposed model demonstrates the highest performance compared to the existing machine-learning algorithms we tested, including classical methods and (graph and non-graph) deep learning. We provide a detailed analysis of each component of our model.
\end{abstract}
\section{Introduction}
Structural connections between brain regions constitute complex brain networks known as the \textit{connectome}. Brain networks are represented by graphs, where each brain region is a node, with edges representing the connections between regions and the edge weight reflecting the strength of the connection. The resulting graph provides a detailed map of brain structural connectivity and can be used to study the organization of brain networks and how they relate to cognitive function and behavior. Structural connectome graphs created from diffusion MRI (dMRI) have been used to study a wide range of neurological and psychiatric disorders, including Alzheimer’s disease (AD) \cite{frau2021conductance,wang2018alterations}, schizophrenia \cite{karlsgodt2010structural,wang2020altered}, and autism spectrum disorders (ASD) \cite{tolan2018graph}, as well as to understand normal brain development and aging \cite{dennis2013development}.

In this line, sex classification using structural brain connectivity has been an important problem \cite{beacher2012autism,williamson2022sex}. Clinically, understanding sex differences in brain connectivity patterns can provide insights into the neurobiology of neurological and psychiatric disorders that have different prevalence rates and symptoms between males and females \cite{gur2016sex}. For example, simple thresholding of the structural brain connectivity has shown different sub-brain networks in males versus females \cite{ingalhalikar2014sex,jahanshad2011sex}. Some studies have found that males and females with ASD \cite{beacher2012autism} and conduct disorder \cite{zhang2014sex} 
have different patterns of structural brain connectivity, which may help to explain differences in the symptomatology of the disorder between the sexes. 
Therefore, predicting sex based on the structural connectome may help to identify potential biomarkers or risk factors for these disorders and to develop more personalized and effective treatments.

The task of classification using the structural connectome involves analyzing high-dimensional and complex data, which can be challenging for traditional statistical approaches. Graph neural networks (GNNs), particularly graph convolutional networks (GCNs) \cite{zhang2019graph}, provide a powerful and flexible framework for analyzing brain connectivity graphs. GCNs can learn from the complex interrelationships between nodes and edges, capturing both local and global patterns in the graph structure. This makes GCNs well-suited for classification and prediction based on structural brain connectivity and for identifying the most predictive brain regions and connections. Furthermore, GCNs can be trained on large datasets, increasing their generalizability and applicability to different populations and contexts.


GCNs can leverage the rich structural information in the connectome to make accurate predictions. They do so by performing iterative message-passing between neighboring nodes in the graph, using learnable functions to aggregate and transform information from neighboring nodes, and updating the features of each node based on the aggregated information. This allows GCNs to capture the complex relationships between brain regions and their connections and make predictions based on this information. A GCN-based model is promising for analyzing structural connectome and has shown great potential for improving our understanding of neurological and psychiatric disorders, as well as normal brain development and aging. 

Several state-of-the-art methods have shown the application of GCNs in sex classification from structural and functional brain networks, achieving high classification accuracy compared to existing machine-learning (ML) methods. In a study combining the GCN model and the long short-term memory (LSTM) network to categorize the functional connectivity of demented and healthy patients \cite{xing2019dynamic}, to enhance the disease classification, gender and age predictions were added to a regularization task. Further, the Siamese GCN has been proposed for metric learning in the context of sex classification \cite{ktena2018metric}.
GCNs are also combined with recurrent neural networks to predict sex on temporal fMRI brain graphs \cite{kazi2022dg}.
The spectral GCN has been employed for the region-of-interest identification in functional connectivity graphs for sex classification as well \cite{arslan2018graph}.

In this paper, we propose a simple yet efficient GCN-based multi-head model capable of differentiating sex using structural brain connectivity. We propose a new design architecture that involves multiple parallel graph/non-graph based operations to collect information from all fronts. Through experiments on two public databases, we show that the proposed model outperforms conventional ML and non-graph-based deep-learning (DL) methods. We continue with a description of the proposed method, experiments, discussion, and conclusion.

%
\begin{figure*}[t]
\begin{center}
\centering
\includegraphics[width=1.0\textwidth]{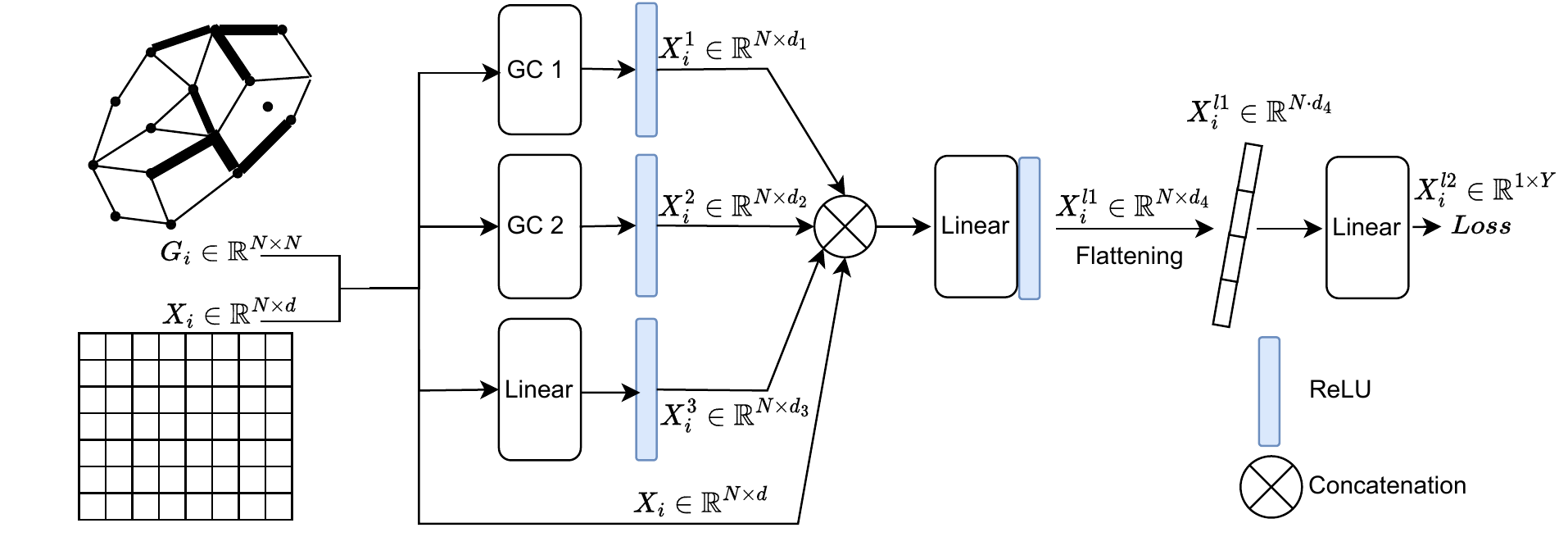}
\end{center}
   \caption{End-to-end pipeline of the proposed model. GC stands for Graph Convolution. The thickness of the edges in $G_i$ shows the weights on the edges. $d, d_1, d_2, d_3,d_4$ are the output dimensions at each layer. $l1$ and $l2$ are the outputs of the two linear layers.}
\label{fig.method}
\end{figure*}

\section{Methods}
Let the dataset be $D =\left( D_1, D_2,..., D_M \right)$ with $M$ subjects. The $i^{th}$ subject is  represented as $D_i \in \left ( G_i , X_i \right )$, i.e., with a brain connectivity graph $G_i \in \mathbb{R}^{N \times N}$ with $N$ nodes, and the corresponding feature matrix $X_i \in \mathbb{R}^{N \times d}$ representing the features for the nodes. The task is to classify each subject $D_i$ into $Y$ classes. We define a model $f_{\theta }$ as: 
\begin{equation}
    y_i = f_{\theta }(G_i, X_i),
\end{equation}
where $\theta$ is the set of learnable parameters. The proposed model $f_{\theta}$ consists of four branches collecting complementary information from the same input setup. 
We employ a combination of GCNConv layers \cite{kipf2016semi}, linear layers, and a skip connection. The GCNConv layers capture low-level features of the graph, while the linear layers learn complex, non-linear relationships between features and make the final classification decision. Lastly, the skip connection helps to address the specific problem of over-smoothing, which may occur when applying graph convolutions. 
GCNConv is based on the graph Laplacian matrix and uses a simple convolutional operation to propagate information between neighboring nodes in the graph. It can be mathematically defined as $X_{i}^{l} = S^{-\frac{1}{2}}G_{i}S^{\frac{1}{2}}X_i\Theta_{l}$ where $S$ is the diagonal degree matrix, $S_{jj} = \sum_{k = 1}^{N}G_{jk}$, and $\Theta_{l}$ is the set of parameters for the $l^{th}$ branch. The linear layers in the model are defined as $X_{i}^{l} = \sigma \left ( \Theta_{l}X_i + b \right )$, where $\sigma$ is the non-linearity (ReLU) function and $b$ is the bias. The motivation for using a combination of GCNConv layers with different embedding sizes is that each one transforms the data into a different space from the same input, hence collecting varied information. The end-to-end pipeline is shown in Figure \ref{fig.method}. We use the weighted cross-entropy loss to train the model. 

\section{Experiments}
We thoroughly analyzed our multi-head GCN model via the task of sex classification with various experiments on two public databases (Table \ref{datasets}). We set aside 10\% of our data. The rest of the data was used for developing and fine-tuneing our models through cross-validation. Here, we first provide details on the datasets, pre-processing and implementation. We then show baseline experiments, followed by a comparison with the state-of-the-art DL-based methods. Further, we show ablation tests for various learning techniques used. Lastly, we show results on the 10\% held-out data to check the model's generalizability.
\begin{table}[t]
\centering
\caption{Description of dataset size (number of available scans), distribution across the classes, and partitioning. Due to missing demographic data, nine subjects were removed from the OASIS3 dataset. The female ratio is the portion of scans from female subjects, providing a baseline prediction accuracy for a constant (always female) predictor.}\label{tab1}
\begin{tabular}{|l|c|c|c|c|c|c|c|}
\hline
Name&Subjects&Total samples&Samples-10\% &10\%&Male &Female&Female ratio\\
\hline
PREVENT--AD&347& 789 &710 &79 &199 &511 &72\% \\
OASIS3 &771&1294 &1164 &121&515 &649 &56\% \\
\hline
\end{tabular}
\label{datasets}
\end{table}
\subsection{Datasets}
\subpara{Pre‑symptomatic Evaluation of Experimental or Novel Treatments for Alzheimer’s Disease (PREVENT-AD) \cite{leoutsakos2016alzheimer}} is a publicly available dataset that aims to provide a comprehensive set of data on individuals who are at risk for developing AD (\url{https://prevent-alzheimer.net}). The database contains neuroimaging studies such as MRI (including dMRI) and PET scans, a range of demographic, clinical, cognitive, and genetic data, as well as data on lifestyle factors such as diet and exercise. The dataset comprises 347 subjects, some with multiple (longitudinal) dMRI scans, totaling 789 dMRI scans.

\subpara{Open Access Series of Imaging Studies, the third release (OASIS3) \cite{lamontagne2019oasis}} is a longitudinal neuroimaging, clinical, and cognitive dataset for normal aging and AD, provided freely to researchers worldwide (\url{http://www.oasis-brains.org}). The OASIS3 dataset contains MRI scans (including dMRI), cognitive assessments, demographic information, and clinical diagnoses for subjects, including healthy controls, individuals with MCI, and AD patients. We used 1294 brain scans from 771 subjects.

\subsection{Pre-processing}
We used FreeSurfer \cite{fischl2012freesurfer} to process the databases (additionally the longitudinal processing pipeline \cite{reuter2012within} for PREVENT-AD). We then ran the FreeSurfer diffusion processing pipeline and propagated the 85 automatically segmented cortical and subcortical regions from the structural to the diffusion space. These 85 regions act as the nodes in our graph setup. Next, we used our public toolbox (\url{http://www.nitrc.org/projects/csaodf-hough}) to reconstruct the diffusion orientation distribution function in constant solid angle \cite{aganj2010reconstruction}, run Hough-transform global probabilistic tractography \cite{aganj2011hough} to generate 10,000 fibers per subject, compute symmetric structural connectivity matrices, and augment the matrices with indirect connections \cite{aganj2014structural}. Once we had all the graphs $G_i$, we performed a population-level normalization on edge weights. For node features, we used the volume, apparent diffusion coefficient, and fractional anisotropy obtained for each region, as well as the row in $G_i$ representing the connectivity to the rest of the brain. Therefore, for each subject we obtained $G_i \in \mathbb{R}^{85 \times 85}$ and corresponding $X_i \in \mathbb{R}^{85 \times 88}$. 

\subpara{Implementation details.} All the experiments were run via 10-fold cross-validation with the same folds across methods and experiments. The data was split into 10 folds based on subjects (rather than scans). For model robustness, we added zero-mean random normal noise with a standard deviation of $0.01$ to the training samples. All the experiments were run on a Linux machine with 512 GB of RAM, an Intel (R) Xeon (R) Gold 6256 CPU @ 3.60 GHz, and an NVIDIA RTX A6000 (48 GB) graphics processing unit. For a fair comparison, we chose the number of layers for comparative methods such that the total numbers of parameters for GCNConv (2453), DGCNN (4653), Graphconv (4653), ResGatedGraphConv (RGGC) (9128), and GINConv (2683) were similar to the proposed method (5073). In our experiments, the values of $d_1$, $d_2$, $d_3$, and $d_4$ were 25, 20, 5, and 2, respectively. We kept 10\% of the data aside from both datasets so as not to heuristically fit the model to the entire data, and tested the model at the end on the unseen data. All the comparative methods are selected from PyTorch geometric \cite{he2022pytorch}.

\begin{table}[t]
\rowcolors{2}{white}{gray!15}
\centering
\caption{Classification results (mean accuracy) using conventional ML methods. \textbf{Bold} and {\color[HTML]{FE0000}red} denote the best and the runner-up, respectively.}\label{tab1}
\begin{tabular}{|l|c|c|}
\hline
Model Type/Dataset &PREVENT-AD &OASIS3 \\
\hline
Tree Coarse Tree & 74.1 & 60.1 \\
Logistic Regression &54.5 &52.6 \\
Naive Bayes (Kernel) &62.0 &55.1 \\
SVM (Quadratic) &\textbf{84.1} &\textbf{72.3} \\
KNN (Weighted KNN) &77.5 &63.6 \\
Ensemble (Boosted KNN) &78.3 &68.0 \\
Ensemble (Subspace Discriminant) &82.8 &64.9 \\
Ensemble (Subspace KNN) &74.2 &59.5 \\
Ensemble (RUSBoosted Trees) &72.5 &67.3 \\
Neural Network (Wide) & {\color[HTML]{FE0000} 83.2} & {\color[HTML]{FE0000} 72.0}\\
\hline
\end{tabular}
\label{baseline}
\end{table}
\subpara{Baselines.} Before testing our model, we applied conventional ML algorithms to determine the baseline performance for the two datasets. We used the Statistical and Machine Learning Toolbox of MATLAB with default parameters. Apart from basic classification methods such as decision trees (Coarse Tree), Logistic Regression, and Kernel-based Naive Bayes, we also tested the Support Vector Machine (SVM) and K-Nearest Neighbors. Table \ref{baseline} presents the performance of these models on both datasets. The results suggest that the SVM (Quadratic) and Neural Network perform best among all the methods. 
\begin{table}[]
\rowcolors{2}{white}{gray!15}
\centering
\caption{Classification results (accuracy mean $\pm$ StD) using DL methods, with and without data augmentation (all with skip connections, but various respective pooling strategies).}\label{comp_methods}
\begin{tabular}{|l|c|c|c|c|}
\hline
 & \multicolumn{2}{c|}{With Augmentation} & \multicolumn{2}{c|}{Without Augmentation} \\ \hline
Model/Dataset & PREVENT-AD & OASIS3 & PREVENT-AD & OASIS3 \\ \hline
MLP & 77.3 $\pm$ 6.5 & 75.1 $\pm$ 4.1 & 77.3 $\pm$ 6.6 & 75.1 $\pm$ 4.0 \\
DGCNN\cite{wang2019dynamic} & 76.8 $\pm$ 5.5 & 75.5 $\pm$ 3.8 & 76.9 $\pm$ 4.8 & 74.4 $\pm$ 4.4 \\
Graphconv\cite{morris2019weisfeiler} & 79.8 $\pm$ 1.3 & 75.2 $\pm$ 4.2 & 80.9 $\pm$ 6.5 & 75.3 $\pm$ 2.9 \\
RGGC\cite{bresson2017residual} & 80.8 $\pm$ 7.2 & 74.8 $\pm$ 4.2 & 80.0 $\pm$ 7.4 & 75.3 $\pm$ 4.3 \\
GINConv\cite{hu2019strategies} & 80.8 $\pm$ 4.2 & 74.3 $\pm$ 3.8 & 80.8 $\pm$ 3.5 & 74.1 $\pm$ 4.4 \\
GCNConv\cite{kipf2016semi} & {\color[HTML]{FE0000} 85.2 $\pm$ 5.8} & {\color[HTML]{FE0000} 81.8 $\pm$ 5.2} & {\color[HTML]{FE0000} 85.9 $\pm$ 5.3} & {\color[HTML]{FE0000} 81.5 $\pm$ 4.4} \\
Proposed & \textbf{90.6 $\pm$ 6.8} & \textbf{88.6 $\pm$ 4.1} & \textbf{89.5 $\pm$ 6.1} & \textbf{87.8 $\pm$ 6.6} \\ \hline
\end{tabular}
\end{table}
\subsection{Results and Discussion}
\subpara{Comparative methods.}
Table \ref{comp_methods} shows the performance of various DL models with and without data augmentation. We show results on augmenting the node features with zero-mean uniform noise with $StD = 0.01$. The table includes the mean and standard deviation of accuracy (across folds) for each model and dataset. The proposed model, with data augmentation, achieves the highest accuracy for both datasets, with GCNConv yielding the second best performance. Overall, augmentation improved the performance and robustness of our model.

\begin{table}[t]
\rowcolors{2}{white}{gray!15}
\centering
\caption{Comparison of the performance (accuracy mean $\pm$ StD) of several GNN models with three different pooling techniques (no augmentation or skip connections).}
\begin{tabular}{|l|c|c|c|c|}
\hline
Dataset & Model/Pooling & Max pool & Mean pool & Flattening \\ \hline
PREVENT-AD & DGCNN\cite{wang2019dynamic} & 77.9 $\pm$ 6.1 & 76.9 $\pm$ 4.8 & 77.9 $\pm$ 6.1 \\
& Graphconv\cite{morris2019weisfeiler} & 80.6 $\pm$ 5.6 & 80.8 $\pm$ 6.5 & 80.9 $\pm$ 5.8 \\
& RGGC\cite{bresson2017residual} & 80.6 $\pm$ 5.9 & 80.0 $\pm$ 7.4 & 81.3 $\pm$ 5.0 \\
& GINConv\cite{hu2019strategies} & 79.9 $\pm$ 6.0 & 80.8 $\pm$ 3.5 & 81.8 $\pm$ 5.0 \\
& GCNConv\cite{kipf2016semi} & \textbf{85.6 $\pm$ 6.1} & {\color[HTML]{FE0000} 85.9 $\pm$ 5.3} & {\color[HTML]{FE0000} 88.3 $\pm$ 5.5} \\
& Proposed & {\color[HTML]{FE0000} 83.6 $\pm$ 4.7} & \textbf{87.5 $\pm$ 5.6} & \textbf{90.5 $\pm$ 5.3} \\ \hline
OASIS3 & DGCNN\cite{wang2019dynamic} & 73.8 $\pm$ 5.2 & 74.7 $\pm$ 3.9 & 74.9 $\pm$ 3.7 \\
& Graphconv\cite{morris2019weisfeiler} & 75.2 $\pm$ 3.8 & 75.0 $\pm$ 3.5 & 75.6 $\pm$ 3.6 \\
& RGGC\cite{bresson2017residual} & 73.0 $\pm$ 3.9 & 76.0 $\pm$ 3.8 & 75.2 $\pm$ 4.6 \\
& GINConv\cite{hu2019strategies} & 73.1 $\pm$ 3.9 & 73.4 $\pm$ 5.2 & 73.5 $\pm$ 4.2 \\
& GCNConv\cite{kipf2016semi} & \textbf{83.5 $\pm$ 4.1} & \textbf{82.5 $\pm$ 4.1} & {\color[HTML]{FE0000} 83.5 $\pm$ 4.2} \\
& Proposed & {\color[HTML]{FE0000} 82.7 $\pm$ 4.1} & \textbf{82.5 $\pm$ 4.1} & \textbf{86.1 $\pm$ 4.5} \\ \hline
\end{tabular}
\label{ablation1}
\end{table}

\subpara{Ablation tests.}
Table \ref{ablation1} shows the accuracy for each model and pooling technique, suggesting that the choice of pooling technique can significantly impact the performance of the models. We use flattening for aggregating the information from the graph nodes, which can be seen to perform best among pooling techniques. Flattening keeps information from the entire graph, i.e. all the nodes and their corresponding learned representation, whereas mean and max pooling smoothes and leaves out the information, respectively. This table highlights the importance of considering different pooling techniques when selecting and evaluating DL models. Table \ref{skip} shows the ablation results on adding a skip connection from the raw input to the final linear layer, suggesting an (insignificant) improvement by the skip connection in most cases. Table \ref{comp_methods} also shows the ablation on augmentation. It can be observed that the results of the proposed method for Prevent-AD dataset is different despite the same setup, this is due to the randomness in initialization of two separate experiments. 
\begin{table}[t]
\rowcolors{2}{white}{gray!15}
\centering
\caption{Experimental results (accuracy mean $\pm$ StD) with and without skip connection (all with flattening and no augmentation).}
\begin{tabular}{|l|c|c|c|c|}
\hline
 & \multicolumn{2}{c|}{PREVENT-AD} & \multicolumn{2}{c|}{OASIS3} \\ \hline
Model/Skip & with skip & without skip & with skip & without skip \\ \hline
MLP &\textbf{74.7 $\pm$ 3.6}&74.6 $\pm$ 4.8&\textbf{74.7 $\pm$ 4.5}&73.9 $\pm$ 4.5\\
DGCNN & \textbf{79.7 $\pm$ 5.1} & 77.9 $\pm$ 6.1 & \textbf{75.3 $\pm$ 4.1} & 74.9 $\pm$ 3.7 \\
Graphconv & \textbf{82.1 $\pm$ 4.6} & 80.9 $\pm$ 5.8 & 75.2 $\pm$ 3.7 & \textbf{75.6 $\pm$ 3.6} \\
RGGG & 80.3 $\pm$ 4.9 &\textbf{81.3 $\pm$ 5.0} & 74.7 $\pm$ 3.6 & \textbf{75.2 $\pm$ 4.6} \\
GINConv & 81.2 $\pm$ 4.3 & \textbf{81.8 $\pm$ 5.0} & 73.2 $\pm$ 3.3 & \textbf{73.5 $\pm$ 4.2} \\
GCNConv & \textbf{88.5 $\pm$ 6.0} & 88.3 $\pm$ 5.5 & \textbf{87.0 $\pm$ 5.6} & 83.5 $\pm$ 4.2 \\
Proposed & \textbf{90.7 $\pm$ 6.3} & 89.03 $\pm$ 5.4 & \textbf{87.4 $\pm$ 6.6} & 86.1 $\pm$ 4.5\\ \hline
\end{tabular}
\label{skip}
\end{table}
\begin{table}
\centering
\caption{Classification accuracy of 10\% held-out (never before seen) data.}
\begin{tabular}{|l|c|c|c|c|c|c|c|}
\hline
Dataset/Method & MLP & DGCNN & Graphconv & RGGG & GINConv & GCNConv & Proposed \\ \hline
PREVENT-AD & 65.6 & 73.1 & 69.9 & 72.0 & 54.8 & {\color[HTML]{FE0000} 75.3} & \textbf{78.5} \\ 
OASIS3 & 69.4 &\textbf{95.9} & 88.4  & 90.9 & 90.1 & 89.3 & {\color[HTML]{FE0000} 95.0}  \\ \hline
\end{tabular}
\label{Unknowndata_exp}
\end{table}

\subpara{Results on held-out data.} Finally, we tested our model on never-before-seen data. Table \ref{Unknowndata_exp} shows results on 10\% of the data that was kept aside before the model development and training. In this experiment, we took the pre-trained model and evaluated the classification accuracy on the new data, which reveals how the model would translate to relatively new data. The flattening technique is applied to the proposed method, whereas maxpooling is applied to comparative methods. It can be observed that the proposed model still performed best for PREVENT-AD and almost tied with DGCNN for OASIS3, showing its superiority with respect to out-of-sample performance.
\section{Conclusion} 
In this paper, we proposed a simple yet effective model capable of capturing complementary information from brain connectivity graphs, which we evaluated in the context of sex classification. The configuration of input data, the initialization of neighborhood information as node features, the combination of GCNConv layers, linear layers, and a skip connection, and eventually the flattening of node features helped to learn better representations of each subject's graph. We have shown that our model outperforms competing techniques on two publicly available datasets, while also ablating several components (augmentation, pooling technique, skip connection). Our results on held-out data further help to measure the model's robustness toward unseen data. In terms of network complexity and size, the proposed model is average-sized, as mentioned in the implementation details. Future work includes the addition of interpretability to the models to find the brain subnetworks responsible for the sex difference, integration of functional and structural connectivity, and evaluation of disease and age prediction. A further step would be to try different graph convolution mechanisms, such as those based on residual connections or gated attention graph convolutions.

\subpara{Clinical Translation.} The proposed method, which takes advantage of GCNs, can be extended from sex classification to clinical prediction and stratification. For instance, it can be used as a biomarker for diagnosis, prognosis, progression/conversion prediction, and treatment effectiveness assessment.
\subsection*{Acknowledgments:} Support for this research was provided by the National Institutes of Health (NIH), specifically the National Institute on Aging (NIA; RF1AG068261).\\ 
Additional support was provided in part by the BRAIN Initiative Cell Census Network grant U01MH117023, the National Institute for Biomedical Imaging and Bioengineering (P41EB015896, R01EB023281, R01EB006758, R21EB018907, R01EB019956, P41EB030006), the NIA (R56AG064027, R01AG064027, R01AG008122, R01AG016495, R01AG070988), the National Institute of Mental Health (R01MH121885, RF1MH123195), the National Institute for Neurological Disorders and Stroke (R01NS0525851, R21NS072652, R01NS070963, R01NS083534, U01NS086625, U24NS10059103, R01NS105820), the NIH Blueprint for Neuroscience Research (U01MH093765), part of the multi-institutional Human Connectome Project, and the Michael J. Fox Foundation for Parkinson’s Research (MJFF-021226). Computational resources were provided through the Massachusetts Life Sciences Center. \\B. Fischl has a financial interest in CorticoMetrics, a company whose medical pursuits focus on brain imaging and measurement technologies. His interests were reviewed and are managed by Massachusetts General Hospital and Mass General Brigham per their conflict-of-interest policies.

%
%
%
%

\bibliographystyle{splncs04}
\bibliography{ref}
\end{document}